\RequirePackage{lineno}
\documentclass[aps,prl,twocolumn,superscriptaddress,showpacs,preprintnumbers,amsmath,amssymb,floatfix]{revtex4}

\usepackage{graphicx}
\usepackage{dcolumn}  

\graphicspath{{ps}}

\begin{document}


\title{ \quad\\[1.0cm] Observation of $\Upsilon(2S)\to\gamma \eta_{b}(1S)$ decay}

\noaffiliation
\affiliation{University of the Basque Country UPV/EHU, 48080 Bilbao}
\affiliation{Beihang University, Beijing 100191}
\affiliation{University of Bonn, 53115 Bonn}
\affiliation{Brookhaven National Laboratory, Upton, New York 11973}
\affiliation{Budker Institute of Nuclear Physics SB RAS, Novosibirsk 630090}
\affiliation{Faculty of Mathematics and Physics, Charles University, 121 16 Prague}
\affiliation{Chonnam National University, Kwangju 660-701}
\affiliation{University of Cincinnati, Cincinnati, Ohio 45221}
\affiliation{Deutsches Elektronen--Synchrotron, 22607 Hamburg}
\affiliation{Duke University, Durham, North Carolina 27708}
\affiliation{Key Laboratory of Nuclear Physics and Ion-beam Application (MOE) and Institute of Modern Physics, Fudan University, Shanghai 200443}
\affiliation{Justus-Liebig-Universit\"at Gie\ss{}en, 35392 Gie\ss{}en}
\affiliation{Gifu University, Gifu 501-1193}
\affiliation{II. Physikalisches Institut, Georg-August-Universit\"at G\"ottingen, 37073 G\"ottingen}
\affiliation{SOKENDAI (The Graduate University for Advanced Studies), Hayama 240-0193}
\affiliation{Gyeongsang National University, Chinju 660-701}
\affiliation{Hanyang University, Seoul 133-791}
\affiliation{University of Hawaii, Honolulu, Hawaii 96822}
\affiliation{High Energy Accelerator Research Organization (KEK), Tsukuba 305-0801}
\affiliation{J-PARC Branch, KEK Theory Center, High Energy Accelerator Research Organization (KEK), Tsukuba 305-0801}
\affiliation{Forschungszentrum J\"{u}lich, 52425 J\"{u}lich}
\affiliation{IKERBASQUE, Basque Foundation for Science, 48013 Bilbao}
\affiliation{Indian Institute of Science Education and Research Mohali, SAS Nagar, 140306}
\affiliation{Indian Institute of Technology Bhubaneswar, Satya Nagar 751007}
\affiliation{Indian Institute of Technology Guwahati, Assam 781039}
\affiliation{Indian Institute of Technology Hyderabad, Telangana 502285}
\affiliation{Indian Institute of Technology Madras, Chennai 600036}
\affiliation{Indiana University, Bloomington, Indiana 47408}
\affiliation{Institute of High Energy Physics, Chinese Academy of Sciences, Beijing 100049}
\affiliation{Institute of High Energy Physics, Vienna 1050}
\affiliation{Institute for High Energy Physics, Protvino 142281}
\affiliation{INFN - Sezione di Napoli, 80126 Napoli}
\affiliation{INFN - Sezione di Torino, 10125 Torino}
\affiliation{Advanced Science Research Center, Japan Atomic Energy Agency, Naka 319-1195}
\affiliation{J. Stefan Institute, 1000 Ljubljana}
\affiliation{Kanagawa University, Yokohama 221-8686}
\affiliation{Institut f\"ur Experimentelle Teilchenphysik, Karlsruher Institut f\"ur Technologie, 76131 Karlsruhe}
\affiliation{Kennesaw State University, Kennesaw, Georgia 30144}
\affiliation{King Abdulaziz City for Science and Technology, Riyadh 11442}
\affiliation{Department of Physics, Faculty of Science, King Abdulaziz University, Jeddah 21589}
\affiliation{Korea Institute of Science and Technology Information, Daejeon 305-806}
\affiliation{Korea University, Seoul 136-713}
\affiliation{Kyoto University, Kyoto 606-8502}
\affiliation{Kyungpook National University, Daegu 702-701}
\affiliation{LAL, Univ. Paris-Sud, CNRS/IN2P3, Universit\'{e} Paris-Saclay, Orsay}
\affiliation{\'Ecole Polytechnique F\'ed\'erale de Lausanne (EPFL), Lausanne 1015}
\affiliation{P.N. Lebedev Physical Institute of the Russian Academy of Sciences, Moscow 119991}
\affiliation{Faculty of Mathematics and Physics, University of Ljubljana, 1000 Ljubljana}
\affiliation{Ludwig Maximilians University, 80539 Munich}
\affiliation{Luther College, Decorah, Iowa 52101}
\affiliation{University of Malaya, 50603 Kuala Lumpur}
\affiliation{University of Maribor, 2000 Maribor}
\affiliation{Max-Planck-Institut f\"ur Physik, 80805 M\"unchen}
\affiliation{School of Physics, University of Melbourne, Victoria 3010}
\affiliation{University of Mississippi, University, Mississippi 38677}
\affiliation{University of Miyazaki, Miyazaki 889-2192}
\affiliation{Moscow Physical Engineering Institute, Moscow 115409}
\affiliation{Moscow Institute of Physics and Technology, Moscow Region 141700}
\affiliation{Graduate School of Science, Nagoya University, Nagoya 464-8602}
\affiliation{Kobayashi-Maskawa Institute, Nagoya University, Nagoya 464-8602}
\affiliation{Universit\`{a} di Napoli Federico II, 80055 Napoli}
\affiliation{Nara Women's University, Nara 630-8506}
\affiliation{National Central University, Chung-li 32054}
\affiliation{National United University, Miao Li 36003}
\affiliation{Department of Physics, National Taiwan University, Taipei 10617}
\affiliation{H. Niewodniczanski Institute of Nuclear Physics, Krakow 31-342}
\affiliation{Nippon Dental University, Niigata 951-8580}
\affiliation{Niigata University, Niigata 950-2181}
\affiliation{Novosibirsk State University, Novosibirsk 630090}
\affiliation{Osaka City University, Osaka 558-8585}
\affiliation{Pacific Northwest National Laboratory, Richland, Washington 99352}
\affiliation{Panjab University, Chandigarh 160014}
\affiliation{Peking University, Beijing 100871}
\affiliation{University of Pittsburgh, Pittsburgh, Pennsylvania 15260}
\affiliation{Theoretical Research Division, Nishina Center, RIKEN, Saitama 351-0198}
\affiliation{University of Science and Technology of China, Hefei 230026}
\affiliation{Showa Pharmaceutical University, Tokyo 194-8543}
\affiliation{Soongsil University, Seoul 156-743}
\affiliation{University of South Carolina, Columbia, South Carolina 29208}
\affiliation{Stefan Meyer Institute for Subatomic Physics, Vienna 1090}
\affiliation{Sungkyunkwan University, Suwon 440-746}
\affiliation{School of Physics, University of Sydney, New South Wales 2006}
\affiliation{Department of Physics, Faculty of Science, University of Tabuk, Tabuk 71451}
\affiliation{Tata Institute of Fundamental Research, Mumbai 400005}
\affiliation{Excellence Cluster Universe, Technische Universit\"at M\"unchen, 85748 Garching}
\affiliation{Department of Physics, Technische Universit\"at M\"unchen, 85748 Garching}
\affiliation{Toho University, Funabashi 274-8510}
\affiliation{Department of Physics, Tohoku University, Sendai 980-8578}
\affiliation{Earthquake Research Institute, University of Tokyo, Tokyo 113-0032}
\affiliation{Department of Physics, University of Tokyo, Tokyo 113-0033}
\affiliation{Tokyo Institute of Technology, Tokyo 152-8550}
\affiliation{Tokyo Metropolitan University, Tokyo 192-0397}
\affiliation{Virginia Polytechnic Institute and State University, Blacksburg, Virginia 24061}
\affiliation{Wayne State University, Detroit, Michigan 48202}
\affiliation{Yamagata University, Yamagata 990-8560}
\affiliation{Yonsei University, Seoul 120-749}
  \author{B.~G.~Fulsom}\affiliation{Pacific Northwest National Laboratory, Richland, Washington 99352} 
  \author{T.~K.~Pedlar}\affiliation{Luther College, Decorah, Iowa 52101} 
  \author{I.~Adachi}\affiliation{High Energy Accelerator Research Organization (KEK), Tsukuba 305-0801}\affiliation{SOKENDAI (The Graduate University for Advanced Studies), Hayama 240-0193} 
  \author{H.~Aihara}\affiliation{Department of Physics, University of Tokyo, Tokyo 113-0033} 
  \author{S.~Al~Said}\affiliation{Department of Physics, Faculty of Science, University of Tabuk, Tabuk 71451}\affiliation{Department of Physics, Faculty of Science, King Abdulaziz University, Jeddah 21589} 
  \author{D.~M.~Asner}\affiliation{Brookhaven National Laboratory, Upton, New York 11973} 
  \author{H.~Atmacan}\affiliation{University of South Carolina, Columbia, South Carolina 29208} 
  \author{V.~Aulchenko}\affiliation{Budker Institute of Nuclear Physics SB RAS, Novosibirsk 630090}\affiliation{Novosibirsk State University, Novosibirsk 630090} 
  \author{T.~Aushev}\affiliation{Moscow Institute of Physics and Technology, Moscow Region 141700} 
  \author{R.~Ayad}\affiliation{Department of Physics, Faculty of Science, University of Tabuk, Tabuk 71451} 
% ...
\collaboration{The Belle Collaboration}

\noaffiliation

\begin{abstract}
We report the observation of $\Upsilon(2S)\to\gamma\eta_{b}(1S)$ decay based on analysis of the inclusive photon spectrum of $24.7$ fb$^{-1}$ of $e^+ e^-$ collisions at the $\Upsilon(2S)$ center-of-mass energy collected with the Belle detector at the KEKB asymmetric-energy $e^+ e^-$ collider.
We measure a branching fraction of $\mathcal{B}(\Upsilon(2S)\to\gamma\eta_{b}(1S))=(6.1^{+0.6+0.9}_{-0.7-0.6})\times 10^{-4}$, and derive an $\eta_{b}(1S)$ mass of $9394.8^{+2.7+4.5}_{-3.1-2.7}$ MeV/$c^{2}$, where the uncertainties are statistical and systematic, respectively.
The significance of our measurement is greater than 7 standard deviations, constituting the first observation of this decay mode.

\end{abstract}


\maketitle

{\renewcommand{\thefootnote}{\fnsymbol{footnote}}}
\setcounter{footnote}{0}

Bottomonium is the system consisting of a $b$ and $\overline{b}$ quark bound by the strong force \cite{qwg}.
The heavy $b$ quark mass allows this system to be described by non-relativistic effective field theory, in addition to phenomenological and lattice methods.
$e^{+}e^{-}$ colliders can directly produce excited bottomonium states, $\Upsilon$, whose radiative decays access the lowest energy spin-singlet bottomonium state, $\eta_{b}(1S)$.
The properties of the ground state are expected to be reliably theoretically calculable.
Study of the $\eta_{b}(1S)$ can further our understanding of the nature of quantum chromodynamics in the non-perturbative regime.

The $\eta_{b}(1S)$ was discovered by the BaBar experiment in $\Upsilon(3S)\to\gamma\eta_{b}(1S)$ decay \cite{babar_y3s}.
Further evidence was provided by BaBar in $\Upsilon(2S)\to\gamma\eta_{b}(1S)$ decay \cite{babar_y2s}, and subsequently by the CLEO experiment \cite{cleo_y3s}.
These analyses studied the inclusive photon spectrum from $\Upsilon$ decays to measure the $\eta_{b}(1S)$ mass ($m_{\eta_{b}(1S)}$) and production branching fractions based on the photon line associated with the hindered M1 radiative transition.
In contrast, subsequent $m_{\eta_{b}(1S)}$ measurements from the Belle experiment have used $h_{b}(nP)\to\gamma\eta_{b}(1S)$ decays produced via $\Upsilon(5S)\to\pi^{+}\pi^{-} h_{b}(nP)$ \cite{belle_y5s} and $\Upsilon(4S)\to\eta h_{b}(1P)$ \cite{belle_y4s}, where $n=1$ and 2.
By measuring the recoil mass against $\pi^{+}\pi^{-}\gamma$, and the mass difference between the $\pi^{+}\pi^{-}$ and $\pi^{+}\pi^{-}\gamma$, and $\eta$ and $\eta\gamma$, recoil masses, the Belle experiment was able to make a complementary measurement of $m_{\eta_{b}(1S)}$.
Other recent measurements have offered compelling but circumstantial information \cite{babar_conversions, cleo_y1s}.

A striking feature of these results is that BaBar and CLEO find $m_{\eta_{b}(1S)} = 9391.1\pm2.9$ MeV/$c^{2}$, whereas Belle measures $9401.6\pm1.7$ MeV/$c^{2}$.
This discrepancy is at the level of 3.1 standard deviations ($\sigma$).
This may be due to experiment-specific systematic effects, or perhaps lineshape distortion in the M1 transition analogous to $J/\psi\to\gamma\eta_{c}(1S)$ \cite{etac_lineshape_ex,etac_lineshape_th}.
There are a large number of $\eta_{b}(nS)$ (where $n=1$ and 2) mass and width predictions from phenomenological quarkonium potential models, non-relativistic QCD, and lattice calculations \cite{burns_summary}.
Theory predictions of the branching fractions vary widely for $\Upsilon(2S)\to\gamma\eta_{b}(1S)$ decays in the range of $(2-20)\times 10^{-4}$ \cite{theory_bf}, and the single experimental measurement is $(3.9\pm1.5)\times 10^{-4}$ \cite{babar_y2s}.
Further $\eta_{b}(1S)$ measurements are necessary for resolving these issues, and reducing the experimental uncertainty in order to discriminate between competing theoretical predictions.

In this Letter, we report a new measurement of $\Upsilon(2S)\to\gamma\eta_{b}(1S)$ decay.
By examining the inclusive photon spectrum, we identify the energy peak associated with this radiative transition, and use it to determine $m_{\eta_{b}(1S)}$ and the branching fraction $\mathcal{B}(\Upsilon(2S)\to\gamma\eta_{b}(1S))$.
This analysis is based on $24.7$ fb$^{-1}$ of $e^+ e^-$ collision data at the $\Upsilon(2S)$ center-of-mass (CM) energy collected with the Belle detector at the KEKB asymmetric-energy $e^+ e^-$ collider \cite{KEKB}.
This data set is equivalent to $(157.8\pm3.6)\times 10^{6}$ $\Upsilon(2S)$ events \cite{ups2s_count}, the largest such sample currently in existence.

The Belle detector is a large-solid-angle magnetic spectrometer consisting of a silicon vertex detector (SVD), a 50-layer central drift chamber (CDC), an array of aerogel threshold Cherenkov counters, a barrel-like arrangement of time-of-flight scintillation counters, and an electromagnetic calorimeter (ECL) comprised of CsI(Tl) crystals. All these are located inside a superconducting solenoid coil that provides a 1.5~T magnetic field.
The ECL is divided into three regions spanning $\theta$, the angle of inclination in the laboratory frame with respect to the direction opposite the $e^{+}$ beam.
The ECL backward endcap, barrel, and forward endcap, cover ranges of $-0.91<\cos{\theta}<-0.65$, $-0.63<\cos{\theta}<0.85$, and $0.85<\cos{\theta}<0.98$, respectively.
An iron flux return located outside of the magnet coil is instrumented with resistive plate chambers to detect $K_L^0$ mesons and muons.
The detector is described in detail elsewhere~\cite{Belle}.
The data collected for this analysis used an inner detector system with a 1.5 cm beampipe, a 4-layer SVD, and a small-cell inner drift chamber.

A set of event selection criteria is chosen to enhance the $\eta_{b}(1S)$ signal while reducing backgrounds from poorly detected photons, $\pi^{0}$ decays, nonresonant production, and other $\Upsilon$ decays.
These criteria are determined by maximizing the figure of merit $S/\sqrt{S+B}$ (where $S$ and $B$ are the number of expected signal and background events, respectively) for each variable under consideration in an iterative fashion.
A subset of ${\sim}5\%$ of the total $\Upsilon(2S)$ data is used as the control sample for optimizing the selection.
To avoid potential bias, these events are discarded from the final analysis.
Large Monte Carlo (MC) samples of simulated $\Upsilon(2S)\to\gamma\eta_{b}(1S)$ events are used as the signal input, assuming the branching fraction from \cite{babar_y2s} and $\eta_{b}(1S)$ decaying to a pair of gluons.
Particle production and decays are simulated using the EVTGEN \cite{evtgen} package, with PHOTOS \cite{photos} for modeling final-state radiation effects, and PYTHIA \cite{pythia} for inclusive $b\overline{b}$ decays.
The interactions of the decay products with the Belle detector are modeled with the GEANT3 \cite{geant} simulation toolkit.

This analysis studies radiative bottomonium transitions based on the energy spectrum of the photons in each event.
Photon candidates are formed from clusters of energy deposited in crystals grouped in the ECL.
Clusters are required to include more than a single crystal.
The ratio of the energy deposited in the innermost $3\times 3$ array of crystals compared to the complete $5\times 5$ array centered on the most energetic crystal is required to be greater than or equal to 0.925.
Clusters must be isolated from the projected path of charged tracks in the CDC, and the associated electromagnetic shower must have a width of less than 6 cm.
Due to increased beam-related backgrounds in the forward endcap region, and insufficient energy resolution in the backward one, we consider only clusters in the ECL barrel region for this analysis, reducing the geometric acceptance by approximately half.

The inclusive photon sample is drawn from events passing a standard Belle definition for hadronic decays.
This requires at least three charged tracks, a visible energy greater than $20\%$ of the CM beam energy ($\sqrt{s}$), and a total energy deposition in the ECL between $0.2\sqrt{s}$ and $0.8\sqrt{s}$.

We consider the cosine of the angle $\theta_{T}$ between the photon and the thrust axis calculated in the $e^{+} e^{-}$ CM frame as a discriminant.
In a given event, the thrust axis is calculated based on all charged particle tracks and photons except the candidate photon.
For continuum background events the photon direction tends to be aligned or anti-aligned along the thrust axis, whereas the distribution for signal events is isotropic.
Therefore to reduce this background we require $|\cos{\theta_{T}}|<0.85$.

To remove backgrounds from $\pi^{0}\to\gamma\gamma$ decays, each photon candidate is sequentially paired with all remaining photon candidates in the event, and vetoed if the resulting invariant mass ($M_{\gamma\gamma}$) is consistent with that of a $\pi^{0}$ ($m_{\pi^{0}}$) \cite{pdg}.
In order to improve purity and reduce combinatorial background, a requirement on the minimum energy of the second photon ($E_{\gamma 2}$) is applied.
We require $E_{\gamma 2}>60$ MeV, and $|M_{\gamma\gamma}-m_{\pi^{0}}|>15$ MeV$/c^{2}$.

\begin{figure}[htb]
  \includegraphics[width=0.95\linewidth]{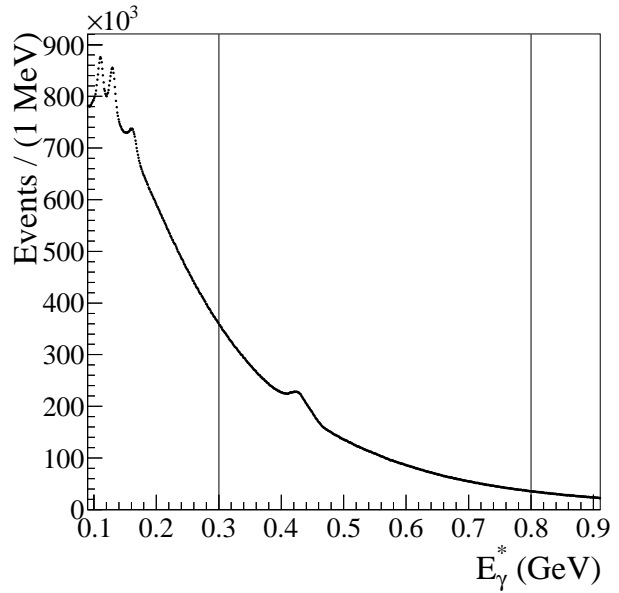}
  \caption{$E_{\gamma}^{*}$ distribution from the data for the photons passing the selection criteria. The visible peaking structures are due to radiative transitions to and from the $\chi_{b0,1,2}(1P)$ states. This analysis is concerned with the $300<E_{\gamma}^{*}<800$ MeV region, indicated by vertical lines. Due to its relative size, an $\Upsilon(2S)\to\gamma\eta_{b}(1S)$ signal expected near 600 MeV is not seen at this scale.}
  \label{fig_all}
\end{figure}

The resulting spectrum of photon energies in the CM frame ($E_{\gamma}^{*}$) is shown in Fig. \ref{fig_all}.
Below 200 MeV there are three prominent peaks related to $\Upsilon(2S)\to\gamma\chi_{b0,1,2}(1P)$ transitions.
The region of interest for this analysis is $300<E_{\gamma}^{*}<800$ MeV, where six components are expected.
Photons from the $\Upsilon(2S)\to\gamma\eta_{b}(1S)$ signal transition will produce a peak in this distribution near $600$ MeV.
Direct production of $\Upsilon(1S)$ via initial-state radiation (ISR), $e^{+} e^{-} \gamma_{ISR} \to \Upsilon(1S)$, results in a second peak at $E_{\gamma}^{*}{\sim}547$ MeV.
A series of three peaks due to $\chi_{bJ=0,1,2}(1P)\to\gamma\Upsilon(1S)$ \cite{footnote_chibJ} transitions are centered at ${\sim}391$, ${\sim}424$, and ${\sim}442$ MeV.
These peaks are Doppler-broadened because the $\chi_{bJ}(1P)$ states originate from $\Upsilon(2S)\to\gamma\chi_{bJ}(1P)$ decays, and are therefore not at rest in the CM frame to which we boost the photon energy for this analysis.
As such, they also overlap one another.
These peaking features are all found above a very large, smooth, inclusive photon background that diminishes as energy increases.

The lineshape parameters and efficiencies are determined from the MC samples.
The $\eta_{b}(1S)$ and $\chi_{bJ}(1P)$ transitions are described by a variation on the Crystal Ball function \cite{cbal}: a bifurcated Gaussian with individual power-law tails on either side.
We assume a natural width for the $\eta_{b}(1S)$ of $\Gamma_{\eta_{b}(1S)}=10^{+5}_{-4}$ MeV \cite{pdg}.
A Gaussian with a low-side power-law tail \cite{cbal} is used to model the ISR-produced $\Upsilon(1S)$ signal.
The underlying background lineshape is parameterized by an exponential function with a sixth-order polynomial.
This was selected based on the best fit of $1.7$ fb$^{-1}$ of continuum background data collected at an energy 30 MeV below the $\Upsilon(2S)$ resonance.

With the above selection criteria our efficiency $(\epsilon)$ for the peaking processes ranges from 26 to 32$\%$, depending on the mode (Table \ref{tab_fit_result}).
The hadronic and photon selections are nearly completely efficient for signal, while the thrust axis and $\pi^{0}$ veto requirements reduce $\epsilon$ by ${\sim}80\%$ and ${\sim}85\%$ respectively.
Photon energy resolution in the CM frame varies from approximately 8 to 12 MeV.
Both quantities increase with energy.

The photon energy scale and resolution are verified with multiple independent control samples.
The Belle $\Upsilon(2S)$ data were collected in two separate time periods with different operating characteristics.
We apply an energy scale adjustment in order to ensure correspondence of the $\chi_{bJ}(1P)\to\gamma\Upsilon(1S)$ transition energies in both of the periods.
To account for differences between MC simulation and data, we fit the energy spectrum with the MC-determined lineshapes for the $\Upsilon(2S)\to\gamma\chi_{bJ}(1P)$ and $\chi_{bJ}(1P)\to\gamma\Upsilon(1S)$ transitions, allowing the energy scale and resolution to vary in order to reproduce the expected $E_{\gamma}^{*}$ values \cite{pdg} of the $\chi_{bJ}(1P)$ peaks in data.
We linearly extrapolate the measured energy scale shift and resolution broadening to the $\eta_{b}(1S)$ energy region, and correct the expected signal lineshape accordingly.

\begin{figure*}[htb]
  \includegraphics[width=0.95\linewidth]{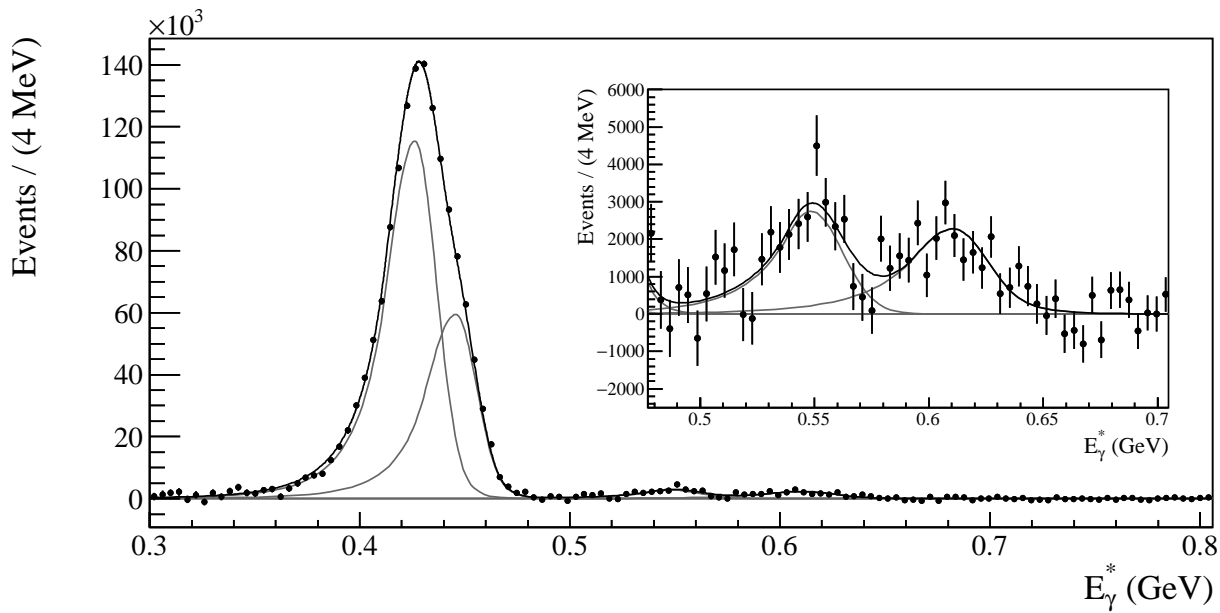}
  \caption{The inclusive photon spectrum after subtraction of the background component of the fit. The black curve indicates the fit to the data, and the gray curves indicate the individual signal components. The $\chi_{b1,2}(1P)\to\gamma\Upsilon(1S)$ transitions at ${\sim}424$ and ${\sim}442$ MeV are dominant. The inset contains the same information with the scale chosen to highlight the ISR and $\eta_{b}(1S)$ signal peaks, appearing at ${\sim}550$ and ${\sim}600$ MeV, respectively.}
  \label{fig_fit_result}
\end{figure*}

We perform a binned maximum-likelihood fit to data in the region of $300<E_{\gamma}^{*}<804$ MeV including all six peaking components and the exponential background.
The yields, energy peak values, and background polynomial coefficients are allowed to vary.
In $\chi_{bJ}(1P)\to\gamma\Upsilon(1S)$ transitions we find the $J=0$ component, known to be suppressed compared to the $J=1$ and 2 transitions, to be absorbed into the other nearby peaks.
We fix the $J=0$ peak position in the fit, and measure a yield consistent with zero.
The results of the fit are shown in Fig. \ref{fig_fit_result} and summarized in Table \ref{tab_fit_result}.
Branching fractions are calculated by dividing the yield by the MC-determined efficiency and number of $\Upsilon(2S)$ events ($(149.6\pm3.4)\times 10^{6}$ with the optimization sample excluded).
The value for $\chi_{bJ}(1P)$ modes includes the $\Upsilon(2S)\to\gamma\chi_{bJ}(1P)$ transition.
The goodness of fit is given by a $\chi^{2}$ per degrees of freedom of $261.5/237$, giving a p-value of 0.132.

\begin{table*}[htb]
\caption{Summary of results. Yield is expressed in thousands of events, with statistical uncertainty only. $\mathcal{B}$ represents the relevant branching fraction, and $E_{\gamma}^{*}$ the corrected transition energy.}
\label{tab_fit_result}
\begin{tabular}
{lcccc}
\hline \hline
Mode & Yield $(10^{3})$ & $\epsilon$ $(\%)$ & $\mathcal{B}$ $(\%)$ & $E_{\gamma}^{*}$ (MeV)\\
\hline
$\chi_{b1}(1P)\to\gamma\Upsilon(1S)$ & $964\pm 8$ & $26.4$ & $2.45{\pm}0.02^{+0.11}_{-0.15}$ & $423.1{\pm}0.1{\pm}0.5$ \\[1mm]
$\chi_{b2}(1P)\to\gamma\Upsilon(1S)$ & $503\pm 6$ & $28.9$ & $1.17{\pm}0.01^{+0.06}_{-0.07}$ & $442.1{\pm}0.2^{+0.5}_{-0.6}$ \\[1mm]
ISR $\Upsilon(1S)$ & $29.2^{+2.9}_{-3.2}$ & $30.0$ & - & $547.2^{+0.6+1.3}_{-2.3-3.2}$\\[1mm]
$\Upsilon(2S)\to\gamma\eta_{b}(1S)$ & $28.8^{+2.6}_{-3.2}$ & $31.6$ & $(6.1^{+0.6+0.9}_{-0.7-0.6}){\times}10^{-2}$ & $606.1^{+2.3+3.6}_{-2.4-3.4}$ \\
\hline \hline
\end{tabular}
\end{table*}

We consider three categories of systematic uncertainties in this analysis: those related to energy calibration, fit parametrization, and all other uncertainties.
These are listed in Table \ref{tab_systematics}, and are summed in quadrature.

As verification of the energy calibration, we consider a complementary method based on the photon energy in the laboratory frame, similar to previous Belle studies \cite{belle_y5s, belle_y4s}.
We derive $E_{\gamma}$-dependent corrections to the photon energy according to the comparison between MC and data for $D^{*0}\to D^{0}(K^{\pm}\pi^{\mp})\gamma$, inclusive $\eta\to\gamma\gamma$, and exclusive $\chi_{b1,2}(1P)\to\gamma\Upsilon(1S)(\mu^{+}\mu^{-})$ decays.
After applying these corrections, only a small remaining resolution broadening, taken as a systematic uncertainty, is required to be applied to the related $E_{\gamma}^{*}$ values to best reproduce the $\chi_{bJ}(1P)\to\gamma\Upsilon(1S)$ transitions in data.
The $\eta_{b}(1S)$ results obtained by these two independent methods agree closely (within 0.2 MeV), providing confidence in our assessment of the energy calibration.

Measurement of the ISR peak position is used to estimate the uncertainty of the $\eta_{b}(1S)$ transition energy.
For this purpose, we adopt the symmetrized combination of the statistical uncertainty from the fit and contributions from the world average $\Upsilon$ mass uncertainties \cite{pdg}.
This value is greater than the maximal difference obtained by repeating the analysis under both energy calibration methods and while varying the derived calibration parameters within $\pm1\sigma$, providing the most conservative bound on this uncertainty.

Alternative parameterizations of the $\eta_{b}(1S)$ transition lineshape are considered by refitting the data using a Breit-Wigner functional form, including the case with additional $E_{\gamma}^{*3}$ corrections suggested for some quarkonium transitions \cite{etac_lineshape_th}.
The latter leads to a $+2.6$ MeV shift in interpretation of the $\eta_{b}(1S)$ transition energy.
The fit is repeated with higher-order $E_{\gamma}^{*}$ contributions considered, but their relative strength cannot be resolved in this anlaysis, and lead to a small additional systematic uncertainty.
We account for uncertainty in the natural $\eta_{b}(1S)$ width by refitting the data according to MC samples generated with the nominal value varied by $\pm1\sigma$ \cite{pdg}.
By comparing $\chi^{2}$ goodness-of-fit results under a variety of different assumed values in this range, we verify that our data are consistent with this nominal value.
We vary the background shape by changing the degree of the polynomial in the exponential to five and seven, and refitting the data.
We also repeat the fit with the background shape fixed to the parameters determined by using only the ISR and $\eta_{b}(1S)$ sidebands: $300<E_{\gamma}^{*}<500$ MeV and $650<E_{\gamma}^{*}<800$ MeV.
The fit is repeated with a $\chi_{b0}(1P)$ yield fixed to the expected value, and the difference in results from its effect on the background shape is taken as a systematic uncertainty.
The systematic effects of fitting with a finer binning of 1 MeV and with an extended range to 900 MeV are also considered.

We assign an overall photon reconstruction efficiency uncertainty of $2.8\%$ based on previous Belle studies of photons in a similar energy range \cite{belle_syst}.
The uncertainty on the number of $\Upsilon(2S)$ events was determined from a study of hadronic decays to be $2.3\%$ \cite{ups2s_count}.
We repeat the measurement of the $\chi_{b1,2}(1P)$ transitions with each selection criterion excluded in turn, and take the difference as the systematic uncertainty related to our modelling of the efficiency.
Derived quantities related to masses and expected CM energies use the world average values and their associated uncertainties \cite{pdg}.

\begin{table*}[htb]
\caption{Summary of systematic uncertainties, divided into those affecting the photon-energy measurement and the overall branching fractions.}
\label{tab_systematics}
\begin{tabular}
  {l|cccc|cccc}
\hline \hline
& \multicolumn{4}{c|}{$E_{\gamma}^{*}$ (MeV)} & \multicolumn{4}{c}{Branching Fraction $(\%)$}\\
Effect & $\chi_{b1}(1P)$ & $\chi_{b2}(1P)$ & ISR & $\eta_{b}(1S)$ & $\chi_{b1}(1P)$ & $\chi_{b2}(1P)$ & ISR & $\eta_{b}(1S)$\\
\hline $E_{\gamma}^{*}$ calibration & ${\pm}0.5$ & ${\pm}0.5$ & $^{+1.2}_{-2.2}$ & ${\pm}2.5$ & $^{+0.1}_{-0.0}$ & $^{+0.1}_{-0.0}$ & $^{+1.9}_{-0.0}$ & $^{+1.1}_{-0.0}$\\[1mm]
$\Gamma_{\eta_{b}(1S)}$ & ${\pm}0.0$ & ${\pm}0.0$ & $^{+0.2}_{-0.0}$ & $\pm0.3$ & $^{+0.2}_{-0.1}$ & $^{+0.0}_{-0.2}$ & $^{+1.1}_{-0.0}$ & $^{+9.9}_{-4.5}$\\[1mm]
Signal shape & $\pm0.0$ & $\pm0.0$ & $^{+0.3}_{-0.4}$ & $^{+2.6}_{-1.0}$ & $^{+0.0}_{-0.1}$ & $^{+0.0}_{-0.1}$ & $^{+1.2}_{-0.2}$ & $^{+10.6}_{-0.1}$\\[1mm]
Background shape & $^{+0.1}_{-0.0}$ & $^{+0.2}_{-0.0}$ & $^{+0.1}_{-2.0}$ & $^{+0.0}_{-2.1}$ & $^{+0.7}_{-0.1}$ & $^{+0.1}_{-0.2}$ & $^{+18.6}_{-1.7}$ & $^{+7.5}_{-2.2}$\\[1mm]
Bin/range & $^{+0.0}_{-0.2}$ & $^{+0.0}_{-0.4}$ & $^{+0.4}_{-0.5}$ & $^{+0.0}_{-0.5}$ & $^{+0.0}_{-1.3}$ & $^{+2.7}_{-0.0}$ & $^{+1.6}_{-0.0}$ & $^{+0.0}_{-4.9}$\\[1mm]
$N(\Upsilon(2S))$& - & - & - & - & ${\pm2.3}$ & ${\pm2.3}$ & ${\pm2.3}$ & ${\pm2.3}$\\[1mm]
$\gamma$ efficiency & - & - & - & - & ${\pm2.8}$ & ${\pm2.8}$ & ${\pm2.8}$ & ${\pm2.8}$\\[1mm]
Selection criteria & - & - & - & - & $^{+2.4}_{-4.8}$ & $^{+2.4}_{-4.8}$ & $^{+2.4}_{-4.8}$ & $^{+2.4}_{-4.8}$\\[1mm] 
\hline Total & ${\pm}0.5$ & $^{+0.5}_{-0.6}$ & $^{+1.3}_{-3.2}$ & $^{+3.6}_{-3.4}$ & $^{+4.4}_{-6.1}$ & $^{+5.1}_{-6.0}$ & $^{+18.7}_{-5.7}$ & $^{+15.3}_{-9.2}$\\
\hline \hline
\end{tabular}
\end{table*}

The corrected peak $E_{\gamma}^{*}$ values of the $\chi_{b1,2}(1P)$ transitions are in good agreement with the world average values (in parentheses) \cite{pdg}: $423.1\pm0.1$ ($423.0\pm0.5$) MeV and $442.1\pm0.2$ ($441.6\pm0.5$) MeV, where the experimental uncertainties are statistical only.
For the $\chi_{b1,2}(1P)\to\gamma\Upsilon(1S)$ branching fractions, we measure $(2.45\pm0.02^{+0.11}_{-0.15})\%$ and $(1.17\pm0.01^{+0.06}_{-0.07})\%$.
These values are consistent with the average of the most recent directly measured values from CLEO \cite{cleo_chib} and BaBar \cite{babar_conversions, babar_chib}: $(2.40\pm0.08)\%$ and $(1.33\pm0.05)\%$.
A significant peak from ISR $\Upsilon(1S)$ events is observed with a corrected $E_{\gamma}^{*}$ value of $547.2^{+0.6+1.3}_{-2.3-3.2}$ MeV, in agreement with the expectation of $547.2\pm0.4$ MeV \cite{pdg}.
The measured ISR signal yield is $(29.2^{+2.9+5.4}_{-3.2-0.9})\times 10^{3}$ events.
This corresponds to the expectation of $(27\pm 3)\times 10^{3}$ events based on the second-order calculation from \cite{isr} and our photon efficiency and ECL angular coverage.

We measure $(28.8^{+2.6+4.2}_{-3.2-2.2})\times 10^{3}$ $\Upsilon(2S)\to\gamma\eta_{b}(1S)$ events, equivalent to a branching fraction of $(6.1^{+0.6+0.9}_{-0.7-0.6})\times 10^{-4}$.
This is in agreement with the most recent lattice QCD calculation of $(5.4\pm1.8)\times 10^{-4}$ \cite{theory_bf}.
This value is compatible with the previous BaBar measurement of $(3.9\pm1.5)\times 10^{-4}$ \cite{babar_y2s}.
We measure a transition energy of $E_{\gamma}^{*}=606.1^{+2.3+3.6}_{-2.4-3.4}$ MeV, to be compared with $609.3^{+5.0}_{-4.9}$ MeV in the similar decay mode in BaBar.
If we consider a transition lineshape proportional to $E_{\gamma}^{*3}$, unlike previous analyses of the M1 radiative transition \cite{babar_y3s,babar_y2s,cleo_y3s}, the interpretation of the data produces a mass measurement of $m_{\eta_{b}(1S)}=9394.8^{+2.7+4.5}_{-3.1-2.7}$ MeV/$c^{2}$.
This is in agreement with the current world average value of $9399.0\pm2.3$ MeV/$c^{2}$ \cite{pdg}.
This is between previous Belle $h_{b}$-based measurements \cite{belle_y5s,belle_y4s} and those from radiative $\Upsilon$ decays \cite{babar_y3s,babar_y2s,cleo_y3s}, consistent with the former at the level of $1.2\sigma$, and $0.7\sigma$ for the latter.
The statistical significance of this measurement is estimated to be $8.4\sigma$, determined from the difference in the likelihood between the results with and without an $\eta_{b}(1S)$ component included.
Even after considering yield-related systematic uncertainties, the signal significance exceeds $7\sigma$.
This result represents the first significant observation of the $\Upsilon(2S)\to\gamma\eta_{b}(1S)$ decay mode.
We look forward to additional dedicated bottomonium data samples from the Belle~II experiment to mitigate energy scale uncertainties and provide greater ability to interpret radiative M1 transition lineshape effects.

We thank the KEKB group for excellent operation of the
accelerator; the KEK cryogenics group for efficient solenoid
operations; and the KEK computer group, the NII, and 
PNNL/EMSL for valuable computing and SINET5 network support.  
We acknowledge support from MEXT, JSPS and Nagoya's TLPRC (Japan);
ARC (Australia); FWF (Austria); NSFC and CCEPP (China); 
MSMT (Czechia); CZF, DFG, EXC153, and VS (Germany);
DST (India); INFN (Italy); 
MOE, MSIP, NRF, RSRI, FLRFAS project and GSDC of KISTI (Korea);
MNiSW and NCN (Poland); MES, 14.W03.31.0026 (Russia); ARRS (Slovenia);
IKERBASQUE and MINECO (Spain); 
SNSF (Switzerland); MOE and MOST (Taiwan); and DOE and NSF (USA).


\begin{thebibliography}{99}
\bibitem{footnote_chibJ}
We use the notation $\chi_{bJ}$ to collectively refer to the $J=0,1,2$ transitions.
\end{thebibliography}
\end{document}